\documentclass[10pt, conference, compsocconf]{IEEEtran}
\usepackage{cite}
\ifCLASSINFOpdf
  \usepackage[pdftex]{graphicx}
\else
  \usepackage[dvips]{graphicx}
\fi

\usepackage[cmex10]{amsmath}
\usepackage{algorithmic}
\usepackage[font=footnotesize]{subfig}
\usepackage{graphicx}

\begin{document}
\title{Investigating Bimodal Clustering in Human Mobility}

\author{\IEEEauthorblockN{James P.~Bagrow }
\IEEEauthorblockA{Center for Complex Network Research\\
and Department of Physics, Northeastern University\\
Boston, MA 02115\\
Email: j.bagrow@neu.edu}
\and
\IEEEauthorblockN{Tal Koren}
\IEEEauthorblockA{Center for Complex Network Research\\
and Department of Physics, Northeastern University\\
Boston, MA 02115\\
Email: t.koren@neu.edu}

}

%


\maketitle

\begin{abstract}
We apply a simple clustering algorithm to a large dataset of cellular telecommunication records, reducing the complexity of mobile phone users' full trajectories and allowing for simple statistics to characterize their properties.  For the case of two clusters, we quantify how clustered human mobility is, how much of a user's spatial dispersion is due to motion between clusters, and how spatially and temporally separated clusters are from one another. 
\end{abstract}


%
\IEEEpeerreviewmaketitle

\section{Introduction}
Recently, much effort has been devoted to understanding, mapping, and modeling
large-scale human and animal trajectories~\cite{BrockmannA,Gonzalez,Viswanathan,Sims, Benichou3,Benichou2,Rhee,Koren,BrockmannB,Fedreschi,Lee,BrockmannC}. Examples include models that describe agents searching a space for a target of unknown position \cite{Benichou2,Benichou3,Koren}; mobility tracking in cellular environments~\cite{Fedreschi,Das,KimB}; human infectious disease dynamics and mobile phone viruses~\cite{BrockmannB,BrockmannC,Lloyd,Kleinberg,Pu}; traffic transportation, and urban planning \cite{Gonzalez}, and references therein. Understanding spatiotemporal patterns and characterizing human mobility and social interactions can now be achieved due to extensive and widespread use of wireless communication devices~\cite{Fedreschi}. 

There is growing experimental evidence that the movement of many species, including humans, can be described by a class of non-trivial random walk models known as L\'{e}vy flights~\cite{BrockmannA,Gonzalez,Klafter}.  A L\'{e}vy flight can be considered a generalization of Brownian motion \cite{Klafter} and belongs to the class of scale-invarient, fractal random processes. L\'{e}vy flights are Markovian stochastic processes in which step lengths $\lambda$ are drawn from a power law distribution: 
\begin{equation}
\lambda(x)\simeq1/|x|^{\alpha+1},
\end{equation}
where $0 < \alpha \leq 2$ is the L\'{e}vy exponent. This implies that the second moment of $\lambda$ diverges and extremely long jumps are possible. For review, see \cite{Chechkin}. 

L\'{e}vy statistics, somewhat controversially, have been found in the search behavior of many species including human hunter-gatherers~\cite{Brown}. L\'{e}vy flight-like movement patterns have been observed while tracking dollar bills  \cite{BrockmannA}, mobile phone users \cite{Gonzalez}, and GPS trajectory traces obtained from taxicabs and volunteers in various outdoor settings \cite{Rhee,Jiang}. 

The mobile phone users studied in \cite{Gonzalez} reveal behavior similar to L\'{e}vy patterns, but individual trajectories show a high degree of temporal and spatial regularity. That investigation analyzed the trajectories of $10^5$ anonymized phone users, randomly selected from more than six million subscribers. The unique mobility patterns found in~\cite{Gonzalez} show a time-independent characteristic travel distance and high probability to frequently return to a few locations. Additionally, real user's mobility patterns may be approximated by  L\'{e}vy flights but only up to a distance characterized by the radius of gyration $r_{g}$. This quantity represents the characteristic distance travelled by a user $a$ observed up to time $t$, defined as
\begin{equation}
r_{g}^{a}(t)=\sqrt{\frac{1}{n_{c}^{a}(t)}\sum_{i=1}^{n_{c}^{a}}(\mathbf{r}_{i}^{a}-\mathbf{r}_{\mathrm{CM}}^{a})^{2}},
\end{equation}
where $\mathbf{r}_{i}^{a}$ represents the $i=1,\ldots,n_{c}^{a}(t)$ positions recorded for user $a$ and $\mathbf{r}_{\mathrm{CM}}^{a}=1/{n_{c}^{a}}(t)\sum_{i=1}^{n_{c}^{a}}\mathbf{r}_{i}^{a}$ is the center of mass of the trajectory~\cite{Gonzalez}. Interestingly, it has been observed that the radius of gyration increases only logarithmically in time, which cannot be explained by traditional L\'{e}vy models; therefore, we must return to the data for further analysis. 
 
\subsection{Motivation and Open Questions}

The work in~\cite{Gonzalez} showed that human mobility patterns are well characterized by the radius of gyration $r_g$.  This is a static measure, in the sense that the order of movement between locations is irrelevant, and can then characterize time-invariant properties of the trajectory.  Mobile phone data only samples the actual underlying trajectory, however, with a user-driven, heterogeneous sampling rate~\cite{Gonzalez}, and this complicates the study of the user's real mobility.  In the same way that $r_g$ avoids these sampling problems by ``integrating'' over time, an appropriate spatial course-graining can provide a basic picture of the time-dependent, evolving characteristics of a subject's mobility pattern. 

In this paper, we apply a simple clustering algorithm to the spatial locations of a user's trajectory.  Finding clusters of frequently visited locations (such as home and work) and collapsing them to a single entity reduces the complexity of the full trajectory while allowing for simple statistics to capture properties relating to how users move between locations.   Interesting questions include:
\begin{enumerate}
\item How spatially separated are such clusters?
\item How often are clusters (re-)visited?  How long do users dwell within clusters?
\item Are larger clusters (more recorded calls over time) more spatially dispersed (as quantified by the $r_g$ of the cluster's elements) than smaller clusters?  How do the $r_g$'s of clusters relate to the total $r_g$?
\end{enumerate}

The remainder of this paper is organized as follows.  We first briefly discuss the dataset (Sec.~\ref{subsec:data}) and the clustering algorithm used to analyze it (Sec.~\ref{subsec:clustering}).  We then introduce several important statistics, calculate them from the dataset at hand, and discuss their implications (Sec.~\ref{sec:results}).  A summary and discussion of future work follows (Sec.~\ref{sec:conc}).

\subsection{Data Set}\label{subsec:data}
As in~\cite{Gonzalez,Pu} we analyze data from a European mobile phone carrier. The data contains the date and time of phone calls and text messages from 6 million anonymous users as well as the spatial location of the phone towers routing these communications. User locations within a tower's service area are not known.  From this full dataset, we select a random subset of 60\,000 users that make or receive at least one phone call during June -- August 2007.  The call history of each user was then used to reconstruct their trajectory of motion during that time period.  

\subsection{Clustering}\label{subsec:clustering}

Our analysis is based on $k$-means clustering which, in general, divides $N$-dimensional populations (or observations) into $k$ distinct sets or clusters by minimizing the intra-cluster sum of squares $w^2$ of an appropriately defined distance metric. Using the notation in \cite{MacQueen} (see also \cite{Gan}):
\begin{equation}
w^2(S)=\sum_{i=1}^{k}\int_{S_i}\left|z-u_i\right|^2 \mathrm{d} p(z),
\end{equation}
where $p$ is the probability mass function for the observations $S=\{S_1,S_2,\ldots,S_k\}$, and $u_i$ $(i=1, \ldots, k)$ is the conditional mean of $p$ over the set $S_i$. For this work, $\mu_i$ will be the center of mass of cluster $i$ and we seek to find the locations of $\mu_i$ that minimize the square Euclidean distance between $\mu_i$ and that cluster's recorded call locations.

For simplicity, we assume $k=2$ clusters throughout this work.  This is a serious assumption, and identifying the correct number of clusters on a per-user basis remains important future work.  However, we have found that the majority of users are well clustered with $k=2$.  To show this, we compute the mean \emph{silhouette value}~\cite{silhouette} for each user.   Define $A(\mathbf{r}_i)$ as the average square (Euclidean) distance from $\mathbf{r}_i$ to all other points in the same cluster, and define $B(\mathbf{r}_i)$ as the average square distance from $\mathbf{r}_i$ to all points in the other cluster.  The silhouette value $s(\mathbf{r}_i)$ for point $\mathbf{r}_i$ is
\begin{equation}
s(\mathbf{r}_i) \equiv \frac{B(\mathbf{r}_i) - A(\mathbf{r}_i)}{\max\left\{B(\mathbf{r}_i), A(\mathbf{r}_i)\right\}},
\end{equation}
and takes values between -1 and 1, with larger values indicating $r_i$ is increasingly well separated from the other cluster.  Taking the average $\left<s\right> \equiv \left<s(\mathbf{r}_i)\right>_i$ over all points then provides a single statistic measuring how well the whole data are clustered.  Poor choices of $k$, for example, lead to smaller $\left<s\right>$~\cite{silhouette}.  We find that 91.8\% of users have $\left<s\right> > 0.8$ and 80.8\% have $\left<s\right> > 0.9$, indicating that the majority of users are well clustered with just $k=2$.

\section{Results}\label{sec:results}
\begin{figure*}[!t]
  \centering
    \begin{minipage}[t]{0.48\linewidth}
    \centering
	\includegraphics[height=\textwidth,angle=-90]{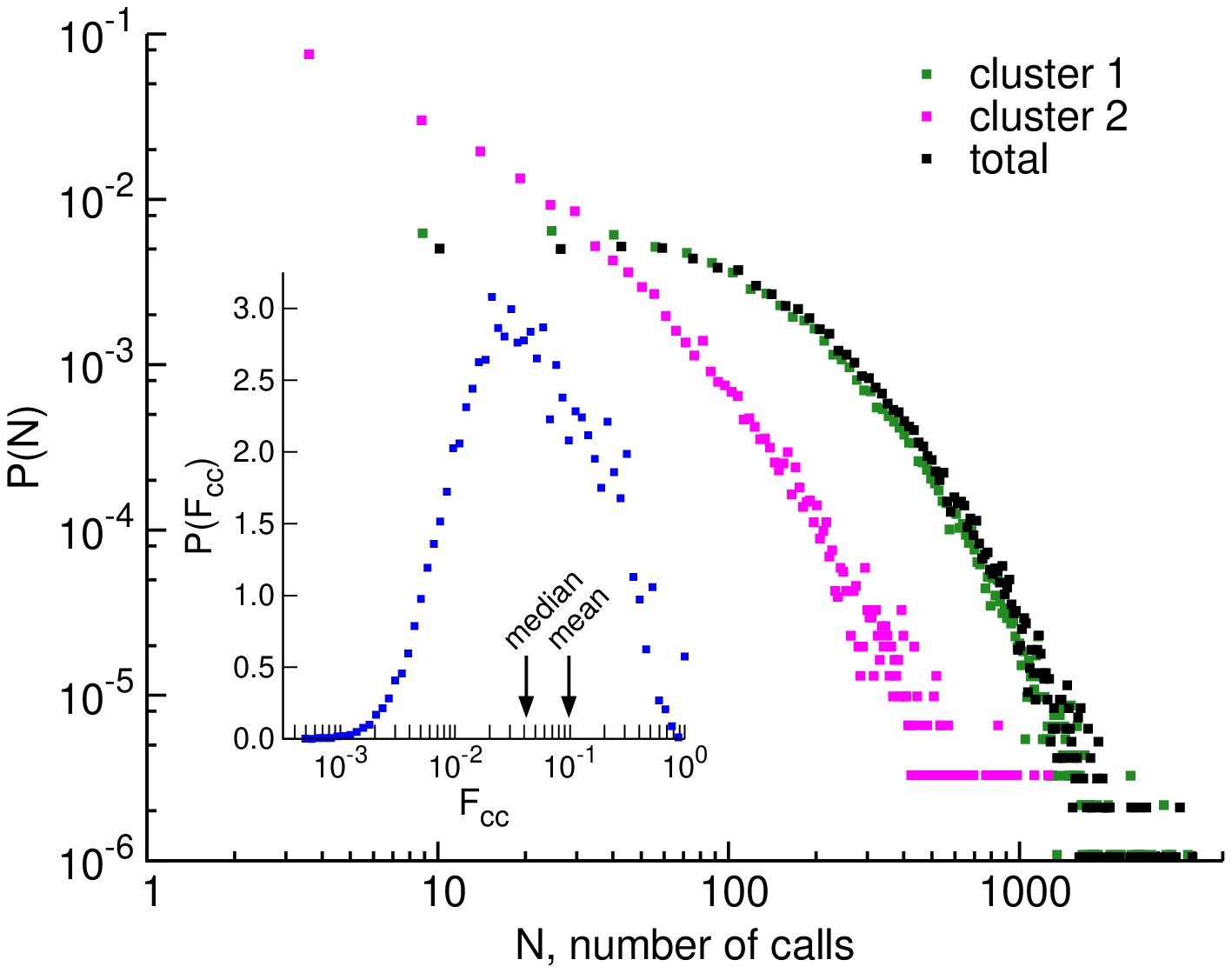}
      \caption{Temporal properties of clusters.  Shown is the distribution $P(N)$ of the number of phone calls  inside each cluster and the total number of calls, $N_1$, $N_2$, and $N_T=N_1+N_2$, respectively.  The majority of calls take place in cluster 1, but a non-negligible amount occur in cluster 2.  (inset) The fraction of jumps $F_\mathrm{CC}$ from one cluster to another, quantifying how often users travel between their clusters.  The primary cluster tends to contain the majority of calls and users tend to move between clusters somewhat rarely, $\left<F_\mathrm{CC}\right>_\mathrm{users}=0.098$, though some move much more frequently.\label{fig:cluster}}
    \end{minipage}
    \hfill
    \begin{minipage}[t]{0.48\linewidth}
    \centering
	\includegraphics[height=\textwidth,angle=-90]{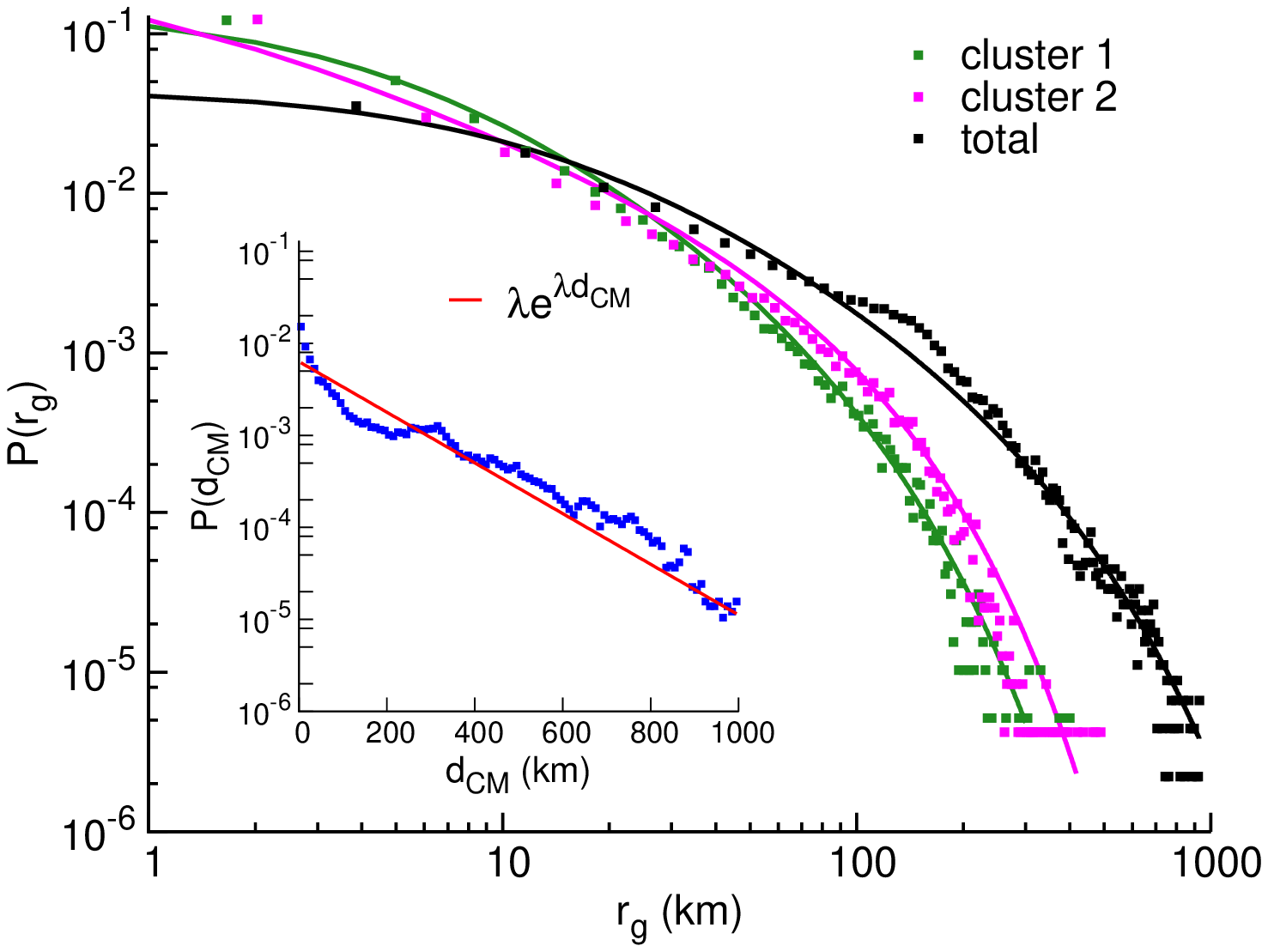}
	\caption{Spread and separation of trajectory clusters.  Shown is the distribution $P(r_g)$ of the radii of gyration for both clusters and over all points, $r_g^{(1)}$, $r_g^{(2)}$, and $r_g^{(T)}$, respectively.  The secondary cluster tends to be slightly more spatially dispersed than the primary cluster. Lines are truncated power laws of the form $\left(r_g + r_g^0\right)^{-\beta_r}e^{-r_g / \kappa}$, characterized by parameters $\Theta \equiv \left(r_g^0,\beta_r, \kappa\right)$.  For the above curves, $\Theta_1 = \left(5.5,1.5,70\right)$, $\Theta_2 = \left(0.75,0.9,70\right)$, and $\Theta_T = \left(15,1.4,260\right)$, for cluster 1, cluster 2, and both, respectively.  (inset) The distribution $P(d_\mathrm{CM})$ of distances between cluster's centers of mass, over all users. The straight line is an exponential distribution with mean $\lambda^{-1} = 157.8$ km, indicating that clusters are often well separated, but distances fall off rather quickly.  \label{fig:spatial}}
    \end{minipage}
\end{figure*}

For each user, we apply the $k$-means algorithm to their trajectory, partitioning the call locations into $k=2$ sets.  
The number of calls in clusters 1 and 2 for user $a$ are $N_1(a)$ and $N_2(a)$, respectively (we identify $N_1(a) \geq N_2(a)$ such that cluster 1 is the primary cluster) and $N_T(a) = N_1(a)+N_2(a)$ is the total number of calls that user $a$ makes during the sample period. The distribution $P(N)$ of the number of calls is shown in Fig.~\ref{fig:cluster}. 

Using the spatial distribution of calls we compute each cluster's center of mass, $\mathbf{r}_{\mathrm{CM}}^{(1)}$ and $\mathbf{r}_{\mathrm{CM}}^{(2)}$, and radius of gyration, $r_g^{(1)}$ and $r_g^{(2)}$, as well as the total center of mass and radius of gyration for all points, $\mathbf{r}_{\mathrm{CM}}^{(T)}$, and $r_g^{(T)}$, respectively.  The distributions of these quantities are shown in Fig.~\ref{fig:spatial}.

To quantify relationships between the two clusters, we compute the separation between their centers of mass, $d_{\mathrm{CM}}$:
\begin{equation}
	d_{\mathrm{CM}} = \left\|\mathbf{r}_{\mathrm{CM}}^{(1)}-\mathbf{r}_{\mathrm{CM}}^{(2)}\right\|,  
\end{equation}
where $\left\|\ldots\right\|$ is the Euclidean norm, and we also count how often a user ``jumps'' between the two clusters.  A user who makes $N_T$ calls will have $N_T-1$ jumps between locations (including remaining at the current location).  We define $F_\mathrm{CC}$ as the fraction of cross-cluster jumps, those that begin and end in different clusters. The distributions of $F_\mathrm{CC}$ and $d_{\mathrm{CM}}$ are shown in the insets of Figs.~\ref{fig:cluster} and \ref{fig:spatial}, respectively.

The number of calls per cluster indicates that users spend the majority of their time in one cluster and visit the other cluster more rarely.  The fraction of cross-cluster jumps is small, $\left<F_\mathrm{CC}\right>_\mathrm{users}$ is less than 0.1, (where $\left<\ldots\right>_\mathrm{users}$ is an average over all sampled users), indicating that the primary cluster provides a stable location in which the user dwells. Likewise, $d_\mathrm{CM}$ is relatively large, $\left<d_\mathrm{CM}\right>_\mathrm{users} = 157.8$ km, indicating that we are finding a semi-frequent but long-distance destination.  It would be interesting to see temporal dependencies on the cluster's occupation probability: are users more likely to be in the secondary cluster on weekends, for example.  

The cluster's radius of gyration summarizes how compact or dispersed user movement is within that cluster.  We find that the larger cluster (in terms of the number of calls) tends to be slightly more spatially compact than the smaller cluster, $r_g^{(1)} < r_g^{(2)}$.  Both $r_g^{(1)}$ and $r_g^{(2)}$ are much smaller than $r_g^{(T)}$, which is to be expected when $d_\mathrm{CM}$ is large.  This means that much of the user's total radius of gyration is generated by movement between two well-separated clusters, as opposed to homogeneous motion over a large space.

\section{Conclusions and Future Work}\label{sec:conc}
We have applied a simple $k$-means clustering algorithm to a large sample of human trajectories generated from mobile phone records.  Doing this characterizes how users move within their set of visited locations and we find that people tend to have one dense, primary cluster and one secondary, dispersed cluster.  Course-graining a user's trajectory into clusters also quantifies how often users move between clusters and we find that users spend the majority of their time in the primary cluster but visit the secondary cluster semi-frequently.  The clusters themselves tend to be well separated, indicating that the secondary cluster is a long-range destination, but the distribution of these distances over all users falls off exponentially quickly, compared to the total radius of gyration.

The most important avenue for future work involves relaxing the assumption of $k=2$ clusters.  While mean silhouette values have shown that the data are well characterized by two clusters, it remains to be seen if introducing more clusters improves the picture.  Furthermore, since so much of a typical user's time is spent in the primary cluster, there remains the tantalizing possibility that further sub-structure is present within it.  In other words, the secondary cluster may represent infrequent long-range trips while the primary cluster may represent the union of home and work clusters, or home and school. Information about important routines such as daily commuting may be contained within the primary cluster.

\section*{Acknowledgment}

The authors would like to thank Marta Gonz\'alez and Albert-L\'aszl\'o Barab\'asi for fruitful discussions, and Pu Wang and Dashun Wang for providing data. This work was supported by the James S. McDonnell Foundation 21st Century Initiative in Studying Complex Systems; NSF within the Dynamic Data Driven Applications Systems (CNS-0540348), Information Technology Research (DMR-0426737), and IIS-0513650 programs; the Defense Threat Reduction Agency Award HDTRA1-08-1-0027; and the U.S. Office of Naval Research Award N00014-07-C. JPB gratefully acknowledges support from DTRA grant BRBAA07-J-2-0035.



%

\end{document}